# Phase instability induced by polar nanoregions in a relaxor ferroelectric system

Guangyong Xu[1], Jinsheng Wen[1,2], C. Stock[3,4], P. M. Gehring[4]

[1]*Condensed Matter Physics and Materials Science Department, Brookhaven National Laboratory, Upton, New York 11973*

[2]*Department of Materials Science, State University of New York, Stony Brook, New York 11794*

[3]*Department of Physics and Astronomy, The Johns Hopkins University, Baltimore, Maryland 21218*

[4]*NIST Center for Neutron Research, National Institute of Standards and Technology, Gaithersburg, Maryland 20899-6100*

**Local inhomogeneities known as polar nanoregions (PNR) play a key role in governing the dielectric properties of relaxor ferroelectrics – a special class of material that exhibits an enormous electromechanical response and is easily polarized with an external field. Using neutron inelastic scattering methods, we show that the PNR can also significantly affect the structural properties of the relaxor ferroelectric $Pb(Zn_{1/3}Nb_{2/3})O_3$-4.5%$PbTiO_3$ (PZN-4.5%PT). A strong interaction is found between the PNR and the propagation of sound waves, i.e. acoustic phonons, the visibility of which can be enhanced with an external electric field. A comparison between acoustic phonons propagating along different directions reveals a large asymmetry in the lattice dynamics that is induced by the PNR. We suggest that a phase instability induced by this PNR-phonon interaction may contribute to the ultrahigh piezoelectric response of this and related relaxor**

**ferroelectric materials. Our results also naturally explain the emergence of the various observed monoclinic phases in these systems.**

Understanding the effects of local inhomogeneities on the properties of materials has always presented a great challenge to condensed matter physicists and materials scientists. The effort to elucidate the role played by local, polar clusters known as polar nanoregions (PNR) that are observed in relaxor ferroelectrics is a prime example of such a challenge. Relaxor ferroelectrics (henceforth "relaxors") belong to a special class of disordered materials that, because of their extraordinary piezoelectric and dielectric properties[1,2,3], show enormous potential for industrial applications such as next generation sensors, actuators, and transducers that convert between mechanical and electrical forms of energy. The inhomogeneities in these materials arise from chemical and valence mixing, and the resulting local polar structures can significantly affect macroscopic properties. For example, one of the defining features of relaxors – the broad and highly frequency dependent dielectric permittivity peak – is directly associated with the relaxation process of the PNR[4], which appear a few hundred degrees above the Curie temperature $T_C$[5]. With cooling the PNR are believed to become larger in volume and/or more ordered in polarity. There have been extensive studies of the interplay between the PNR and long-range polar order in relaxor systems[6,7,8]. On the other hand, the relationship between the PNR and the electromechanical and other structural properties of relaxor systems remains unclear.

The structures of the PNR have been well characterized in many lead based, perovskite relaxor systems[9,10,11,12,13,14] and in particular in the prototypical relaxors $Pb(Zn_{1/3}Nb_{2/3})O_3$ (PZN), $Pb(Mg_{1/3}Nb_{2/3})O_3$ (PMN), and their solid solutions with the conventional ferroelectric $PbTiO_3$ (PT). The local structure of PNR is usually different from the average lattice structure of the compound in which they reside. For example, whereas the PNR in PMN-$x$PT and PZN-$x$PT (for small $x$) exhibit orthorhombic <110> polarizations[14], the average macroscopic structure of these compounds is actually rhombohedral (R). In this article, we establish a link between these different structures by studying the influence of the PNR on the lattice dynamics in a relaxor system. In general, atoms in solids can move about their equilibrium positions and, through their motion, propagate energy in the form of sound waves, or acoustic phonons. We find that the dynamics of relaxors is significantly influenced by these local inhomogeneities. Specifically, we have discovered a strong coupling between the PNR and transverse acoustic (TA) phonons polarized along <110>. The PNR can scatter TA phonons having a parallel polarization; such phonons thus propagate more slowly and have shorter lifetimes. In other words, although the lattice of the bulk still remains virtually cubic (having only a slight rhombohedral distortion), the material itself can appear "softer" along a particular <110> direction due to the effects of the PNR. Our work therefore provides a scenario wherein the interaction of the PNR and the bulk lattice introduces an underlying structural instability, which could provide the microscopic origin of the large piezoelectric properties of relaxor systems.

The system we have studied is $Pb(Zn_{1/3}Nb_{2/3})O_3$-4.5%$PbTiO_3$ (PZN-4.5%PT), which lies on the left (rhombohedral) side of the morphotropic phase boundary (MPB) in the region of the PZN-$x$PT phase diagram where the ultra-high piezoelectric response is observed; indeed, the piezoelectric coefficient $d_{33}$ of this crystal is one of the highest known. While a monoclinic phase can be induced in PZN-4.5%PT by applying an electric field along the [001] direction[15], under zero field it still exhibits a cubic (C) high-temperature phase and a rhombohedral (R) low-temperature phase, with a Curie temperature $T_C \sim 475$ K. Neutron inelastic scattering measurements were performed in the neighborhood of the (220) and ($\bar{2}$20) Bragg peaks (expressed using pseudocubic notation) using the constant-$Q$ method, i.e. by scanning the neutron energy transfer $\hbar\omega$ while sitting at a fixed wavevector transfer $Q$ located a distance $q$ in reciprocal space from a given Bragg peak $G$ ($Q = G + q$). It is important to note that both the PNR and phonon neutron scattering cross sections involve the atomic displacements, whether static or dynamic; therefore both cross sections are subject to the factor $|Q \bullet \varepsilon|^2$, where $\varepsilon$ is the unit vector along the polarization (atomic displacement) direction. Consequently, near the (220) (($\bar{2}$20)) Bragg peak neutron scattering probes only those PNR with [110] ([$\bar{1}$10]) polarizations as well as [110] ([$\bar{1}$10]) polarized phonons propagating along $q$. This situation is shown schematically in Fig. 1A, where $q$ is chosen transverse to the Bragg peak $G$ such that we measure TA phonons propagating along [110] and [1$\bar{1}$0] near ($\bar{2}$20) and (220), respectively.

When PZN-4.5%PT is zero-field cooled (ZFC), the diffuse scattering from PNR forms ellipsoids[13, 14] of equivalent shape and intensity centered on the (220) and ($\bar{2}$20) Bragg peaks. In order to identify the interaction between the PNR and phonons explicitly, an ability to tune the diffuse scattering intensities independently and watch for changes in the phonon spectra is required. Fortunately, this can be achieved by cooling the single crystal sample under a moderate external electric field E = 2 kV/cm applied along [111]. When this is done, PNR with different polarizations are reoriented, which leads to a redistribution of the diffuse scattering in reciprocal space[16] as shown in Fig. 1A. A linear scan of the scattering intensity measured along [H, 2.1, 0] after field cooling to 200 K (Fig. 1B) includes contributions from PNR having both [110] and [$\bar{1}$10] polarizations and illustrates this redistribution. In other words, diffuse scattering intensities from PNR having [110] polarizations (shown in blue in Fig. 1 A and B) weaken while those from PNR having [$\bar{1}$10] polarizations (shown in red in Fig. 1 A and B) strengthen when the system is field-cooled (FC) under a field applied along [111]. Remarkably, we find that the TA phonons polarized along [110] and [$\bar{1}$10] are significantly modified at the same time. Figs. 1 C and D show contour maps of the phonon intensities measured near the (220) and ($\bar{2}$20) Bragg peaks after field cooling to T=200 K. Near (220), the lower-energy acoustic mode (i.e. the TA2 mode polarized along [110]) is sharp and well-defined in energy, whereas the identical TA2 mode (but polarized along [$\bar{1}$10]) near ($\bar{2}$20) is comparatively very soft and broad. By contrast, the transverse optic (TO2) modes are essentially unaffected by the diffuse scattering as they do not differ noticeably between the two Bragg peaks.

Constant-$Q$ scans are shown in Fig. 2 to provide a comparison between FC and ZFC results at different temperatures. After cooling below T$_C$, a difference becomes noticeable between the ZFC (black solid lines) and FC spectra measured near (220) (blue lines) and ($\bar{2}$20) (red lines). At 400 K, the data in Fig. 2 B and E show that where the diffuse scattering is strong (near ($\bar{2}$20)), the corresponding TA2 phonon is soft and broad in energy; conversely, where the diffuse scattering is weak (near (220)), the TA2 mode is hard and well-defined. These results unambiguously demonstrate the presence of a strong coupling between the diffuse scattering (PNR) and the TA2 phonons in PZN-4%PT, evidence of which has also observed in the relaxor PMN[17].   If the PNR grow upon cooling, then this effect should become more pronounced at lower temperatures; the data in Figs. 2 C and F, measured at 200 K, clearly show this to be the case.

It is important to note that if we average the two FC phonon spectra measured below $T_C$ near (220) and $(\bar{2}20)$ (represented by the black lines in Fig. 2), the result agrees almost exactly with the ZFC spectra. Since our ZFC measurements were all performed by heating the sample well above $T_C$ and then cooling in zero field, any residue electric field effect or field-induced piezoelectric strain should have been completely removed. This shows that cooling under this moderate field merely rearranges the multiple <111>-polarized ferroelectric domains such that the system adopts a single domain state with polarization along the [111] field direction; any other effect of the field on the phonons or PNR is negligible. In other words, the electric field does not soften or broaden the TA phonon modes, and the PNR-phonon coupling effect is intrinsic.   If this were not true, then the average of the FC spectra would not match the ZFC spectra.   In the ZFC state the effects of the PNR on the phonon modes are still present (for $T < T_C$); however because the ZFC data represent an average over different domains, the measurements made near (220) and $(\bar{2}20)$ cannot be distinguished.

The half widths at half maximum (HWHM) in energy of the TA2 phonons measured at 200 K are shown in Fig. 3 C. The broad widths of the $[\bar{1}10]$ polarized TA2 phonons indicate a short-lived mode. While cooling in a field oriented along [111] stabilizes a single domain R phase, many PNR having $[\bar{1}10]$ polarizations are present[16]; these PNR appear to interact strongly with $[\bar{1}10]$ polarized TA2 phonons, thereby reducing the phonon lifetime. However, because there are far fewer PNR having [110] polarizations, the [110] polarized TA2 phonons are relatively unaffected (see Fig. 3 A and C).

The data in Fig. 3 A show that the TO2 mode is overdamped at small $q$ at 600 K, in agreement with previous measurements on similar systems[18,19,20]; this q-dependent damping produces the dynamical feature known as the "waterfall". Upon cooling to 200 K, the optic modes harden and sharpen in energy (see Fig. 3 A and B), and are unaffected by the diffuse scattering changes below $T_C$ as can be seen by comparing the TO2 modes near the two Bragg peaks.  This behaviour differs from the strong TA-TO coupling effects previously observed in similar systems at temperatures above $T_C$[20,21]. For $T < T_C$ the lack of any significant change in the TO modes compared to the large change in TA mode energies (between the two Bragg peaks) proves that any coupling between the TA and TO modes with <110> polarizations is extremely weak.

Phonons describe collective modes of atomic motions; acoustic phonons are directly associated with strains in the crystal lattice. A very soft and overdamped acoustic phonon mode is usually indicative of a phase instability and a tendency toward a structural transition. After the phase transition takes place and the system becomes stable, the phonon mode recovers and becomes well-defined again. For instance, the softening of acoustic modes is common in other ferroelectric systems where the TA-TO coupling is strong, and the soft TA mode recovers after the structural phase transition takes place[22]. However, in PZN-4.5%PT the structure remains rhombohedral for $T < T_C$. While no phase transition associated with this $[\bar{1}10]$-TA2 phonon anomaly ever occurs, the phonon mode remains anomalously soft and broad, and the system remains (structurally) unstable at low temperature. In other words, the anomaly in this $[\bar{1}10]$-TA2 mode is indicative of a phase instability in the low temperature R-phase of PZN-4.5%PT.

Despite the large orthorhombic-type asymmetry between the TA2 phonon modes polarized along [110] and $[\bar{1}10]$, the static strain, or the deviation from a cubic structure, along these two orthorhombic directions in PZN-4.5%PT is rather small. The average structure of PZN-4.5%PT at low temperature is rhombohedral with a relatively small distortion with or without a moderate electric field applied along [111]. In our case, the static strain $\Delta d / d \leq 0.2\%$ along the [110] direction (determined from the size of the splitting between Bragg peaks - see also ref. 23); this is clearly not consistent with the large (>70%) difference in phonon energies between the two TA2 modes polarized along [110] and $[\bar{1}10]$, especially when compared with values observed in other relaxor/ferroelectric systems having similar structures. For example, in PMN-60%PT[24] at 400 K in the tetragonal phase, one has a much larger strain $\Delta d / d \sim 1.7\%$ along <001> while the splitting of the TA1 mode is less than 30% (estimated from the ZFC TA1 phonon broadening). Another example is the classic ferroelectric $PbTiO_3$, for which the tetragonal strain is about 6% at 200 K, whereas the TA1 phonons along the $a$ and $c$ directions only differ by ~20% in energy[25]. Note that these two systems have relatively large static strains, are both ferroelectric, and show no evidence of PNR. Intriguingly, similar to what we observe in PZN-4.5%PT, phonon studies on another relaxor system where PNR are present, $K_{0.965}Li_{0.035}TaO_3$ (KLT $x$=0.035)[26], report a large splitting ($\Delta\omega$~25% at T=4.8 K) of the TA1 phonon modes polarized along the $a$ and $c$ directions as well, while the static tetragonal strain in that system is small (~0.2%). These examples suggest that in ferroelectric perovskites, the situation where a small static strain is accompanied by a large asymmetry in the lattice dynamics is unique to relaxors where PNR are present. Once the external field and small static lattice strain have been excluded as origins of the anomalous asymmetry between TA modes measured at $(\bar{2}20)$ and $(220)$, an interaction between the TA modes and the PNR becomes the only viable explanation.

For materials to be considered "good" piezoelectrics it is not enough to exhibit a large static strain (e.g. $PbTiO_3$ is not considered a good piezoelectric despite its large tetragonal strain); instead a material must have a large derivative of strain with respect to external electric field, i.e. the lattice/strain must be able to change significantly under field. Various previous theoretical, computational, and experimental work[27,28,29,30,31] has suggested a link between the phase instability in these compounds and the enhanced piezoelectric responses. In fact, acoustic phonons are directly related to the elastic constants in solids, and a soft TA mode is consistent with a system that is easier to distort (along a certain direction). Our results provide strong evidence that in relaxors such a phase instability does exist and is indeed induced by the PNR and manifested as an asymmetry in the lattice dynamics. Note that this instability is intrinsic and not dependent on the external field (along [111]); the [111] field only rearranges large ferroelectric domains for the effect to be observed macroscopically as discussed previously. In the case of an external field oriented along [011] or [001], where the field direction does not align with the natural rhombohedral ([111]) domain polarization, the intrinsic instability apparently helps to facilitate the field-induced structural change and contributes to the large piezoelectric response[32]. Nevertheless, how or if the PNR react to the field is not yet entirely clear; only very limited studies exist so far, e.g. Ref. 33, and further study is required.

While we have explicitly shown the existence of a strong PNR-phonon interaction in PZN-4.5%PT, this effect is not limited to one composition. It has been shown by neutron and x-ray diffuse scattering measurements that PNR are present in all compositions to the left of the MPB in both PMN-$x$PT and PZN-$x$PT solid solutions[14,34] as well as for compositions inside the MPB (see Fig. S1). There is also evidence for a PNR-phonon interaction in other PT compositions such as PMN-20%PT[35] and pure PMN[17] where an anomalous TA phonon broadening has been observed and attributed to diffuse-phonon coupling. While PT doping toward the MPB or an external electric field[31] does enhance the electromechanical and piezoelectric properties in both PMN-$x$PT and PZN-$x$PT, the relaxor character itself plays an important and fundamental role. In fact, even in the absence of any PT, both pure PZN and PMN already exhibit extraordinarily high field induced piezoelectric strains[36]. With increasing PT content some relaxor properties, such as the frequency dispersion in dielectric permittivity, are gradually suppressed; yet the elastic diffuse scattering from PNR is still present near the MPB. In fact, diffuse scattering measurements from PMN-$x$PT single crystals[34] have suggested that the $\mathbf{Q}$-integrated diffuse scattering intensity increases with increasing PT and reaches a maximum near the MPB – precisely where the piezoelectric response is optimal, which is likely more than just coincidence. When these relaxor-ferroelectric solid solutions cross the MPB into the tetragonal side of the phase diagram with even larger PT content, PNR are no longer present[24,34] and the piezoelectric property drops dramatically[1] (to values below even those of pure PMN and PZN).

The orthorhombic asymmetry in the lattice dynamics induced by the PNR also has clear implications on the emergence of various monoclinic (M) phases in relaxor ferroelectric solid-solutions. Monoclinic phases[37,38,39,40]   have been found in PMN-xPT and PZN-xPT for compositions near the MPB and can be induced in lower PT compositions with a external field applied along [001][41,42]. These are believed to appear as a result of the "polarization rotation" scheme proposed by Fu and Cohen[27]. Here, the polarization of the system can be rotated by an external field in a monoclinic plane, instead of being confined to a certain crystallographic orientation as in the R or T phases (see Fig. 4 for details). While PT doping tends to drive the system toward a tetragonal (T) phase due to the large tetragonal strain of $PbTiO_3$ itself, the low symmetry M phases are considered as a bridge between the R and T phases. One dilemma remains however. Intuitively, the bridging phase for low-PT R and high-PT T phases should be $M_A$, where the polarization lies in the (110) plane (see Fig. 4B); in reality, however, in both PZN-$x$PT and PMN-$x$PT systems the zero-field phase near the MPB is $M_C$, not $M_A$ (in fact it is orthorhombic (O) in PZN-$x$PT, which is a special case of $M_C$), where the polarization can rotate in the (001) plane (Fig. 4C). An $M_C$ phase is also often observed when an electric field is applied along [001]. Using our results on the orthorhombic-type strain induced by PNR, manifested by an asymmetry in the lattice dynamics, this situation can now be understood in simple phenomenological terms. The combination of a rhombohedral (R) distortion and a tetragonal strain (T), either from PT doping or an external [001] field, can only produce an $M_A$ phase; however both $M_A$ and $M_C$ phases are possible (see Fig. 4 B, C) by combining T distortion with an orthorhombic (O) strain instead.

In summary, we have shown that PNR significantly affect the lattice stability in relaxor systems through a fundamental interaction with transverse acoustic phonons. While the average, static structure of the bulk is not explicitly modified by the PNR, the local structure of the PNR can be mapped onto the low energy lattice dynamics macroscopically. The resulting relaxor phase has a nearly cubic structure with a large asymmetry in the lattice dynamics that also helps to explain intuitively the emergence of various M phases in relaxor ferroelectric solid solutions. Our results therefore provide evidence that the ultra-high piezoelectric properties of relaxor ferroelectrics may arise from a structural instability resulting from a competition between the static bulk structure and local inhomogeneities that is mediated by acoustic phonons.

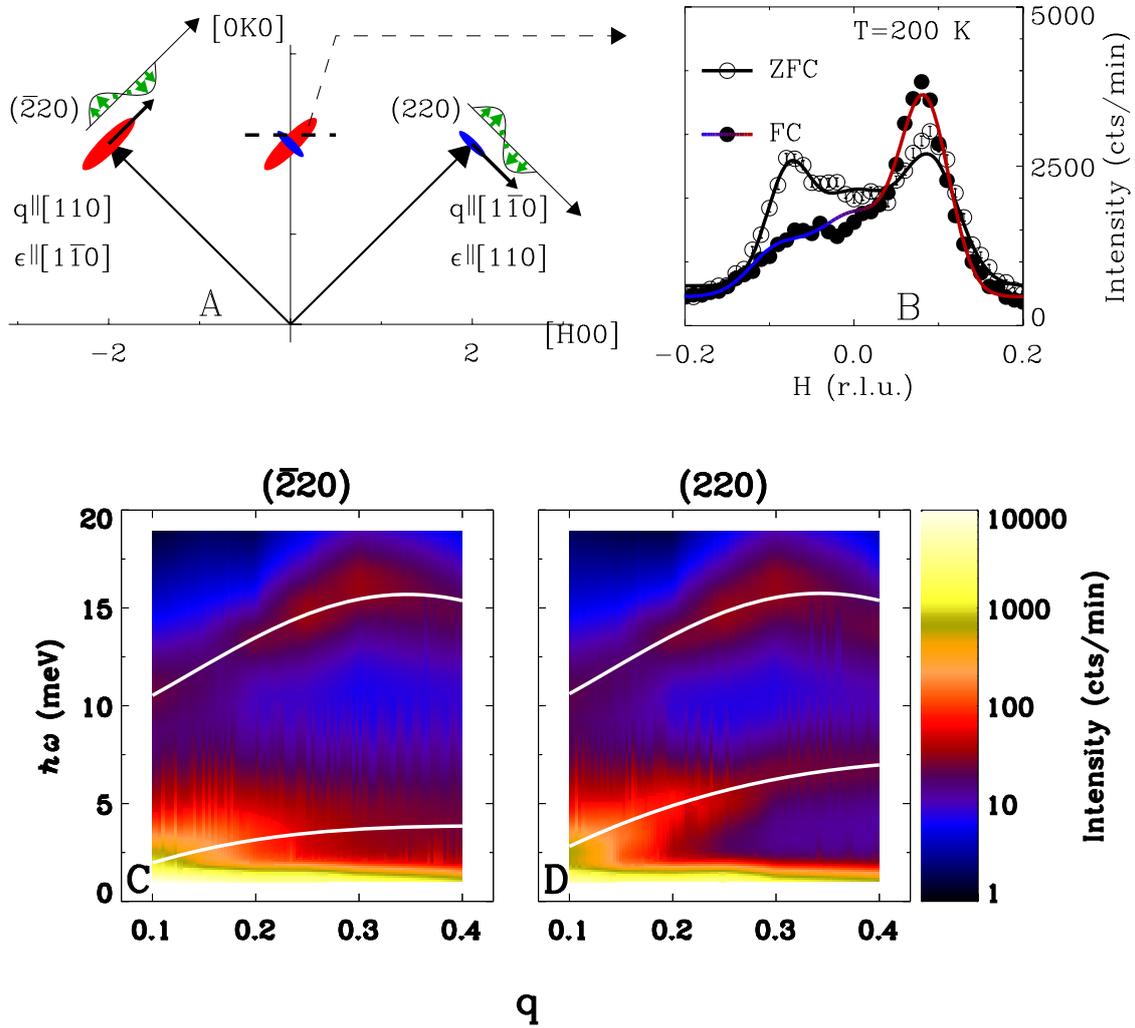

**Figure 1. Neutron scattering measurements performed on a PZN-4.5PT single crystal with dimensions of 10×10×3 mm³.** The experiment was performed on the NCNR BT7 triple-axis-spectrometer using beam collimations of 50'-50'-40'-240'. The final neutron energy was fixed at 14.7 meV. A pyrolytic graphite (PG) filter was placed after the sample to reduce higher order neutrons. Lines are guides to the eye. (A) A

schematic diagram of the neutron scattering measurements, performed near the (220) and $(\bar{2}20)$ Bragg peaks. The blue and red ellipsoids represent the FC diffuse scattering intensity distributions for E along [111]. The polarization and propagation vectors for the phonons are also noted. (B) Profile of the diffuse scattering intensity measured along (H 2.1 0) (dashed line in (A)) under ZFC and FC conditions. (C) Intensity contours measured near $(\bar{2}20)$. (D) Intensity contours measured near (220).

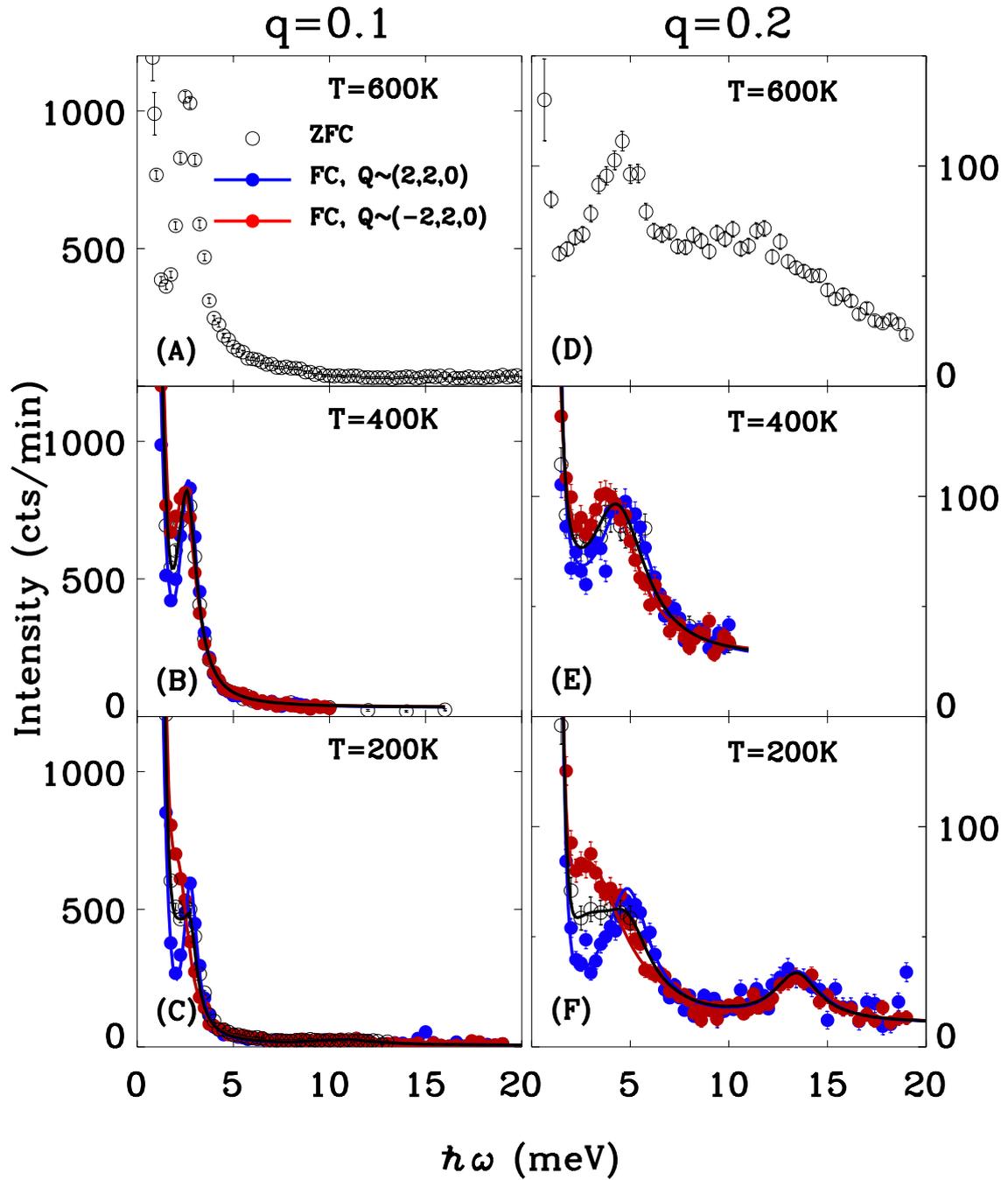

**Figure 2. Constant-Q scans measured near the (220) and $(\bar{2}20)$ Bragg peaks at q=0.1 and 0.2 r.l.u.** The blue and red points are data taken near $(\bar{2}20)$ and (220) after cooling in a field E=2 kV/cm oriented along [111]. The black points are ZFC data. Red

and blue lines are guides to the eye. The black lines are calculated as an average of the two FC data sets (blue and red data points), and can be compared directly to the ZFC data.

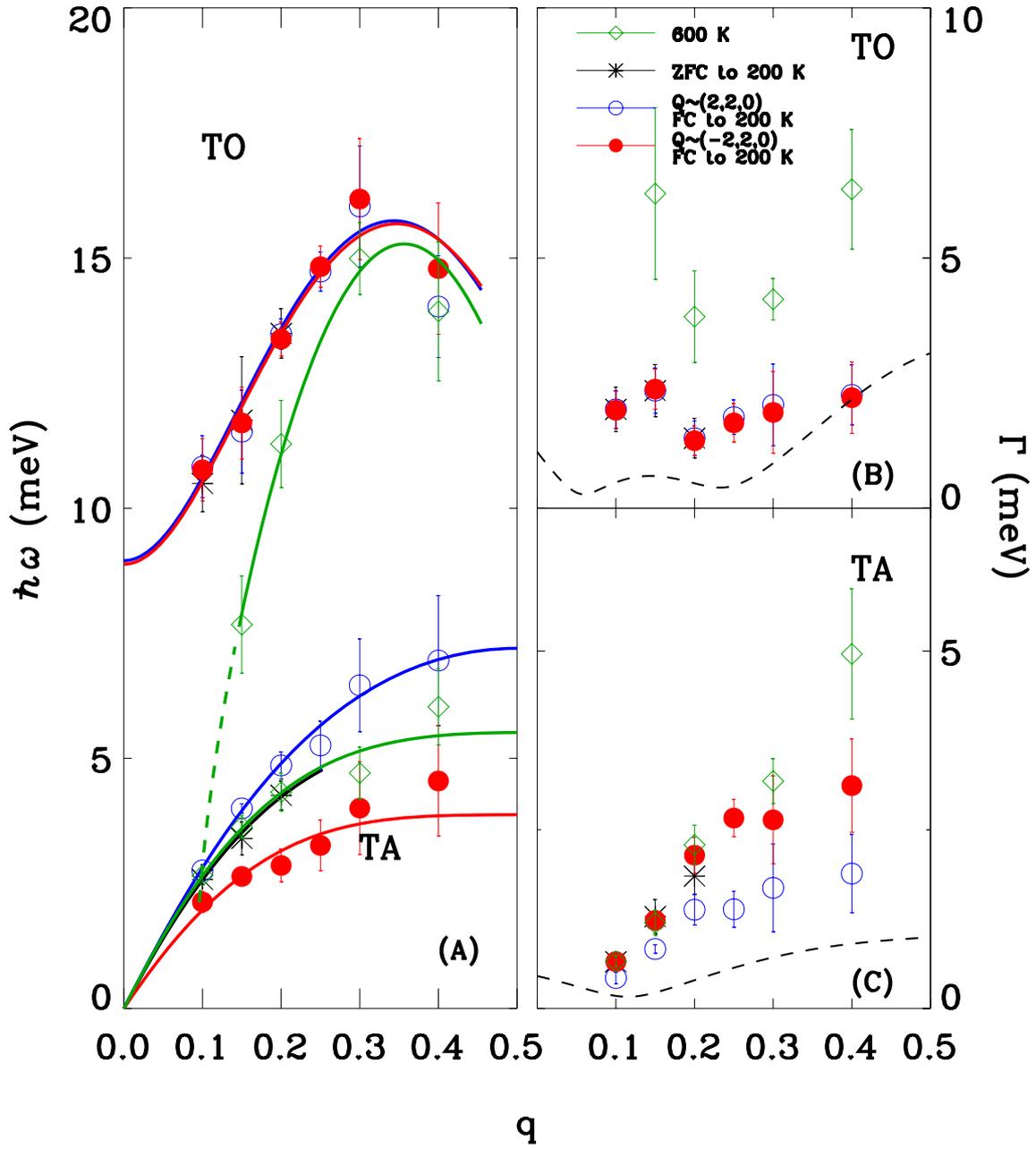

**Figure 3. Phonon dispersions and energy widths (HWHM) measured around the (220) and ($\bar{2}$20) Bragg peaks under FC and ZFC conditions.** The solid lines in (A) are guides to the eye. The dashed lines in (B) and (C) denote the instrumental energy

resolution. The error bars in the figure are fitting errors obtained though least-square fits to the data, assuming the phonon modes having Lorentzian line shapes.

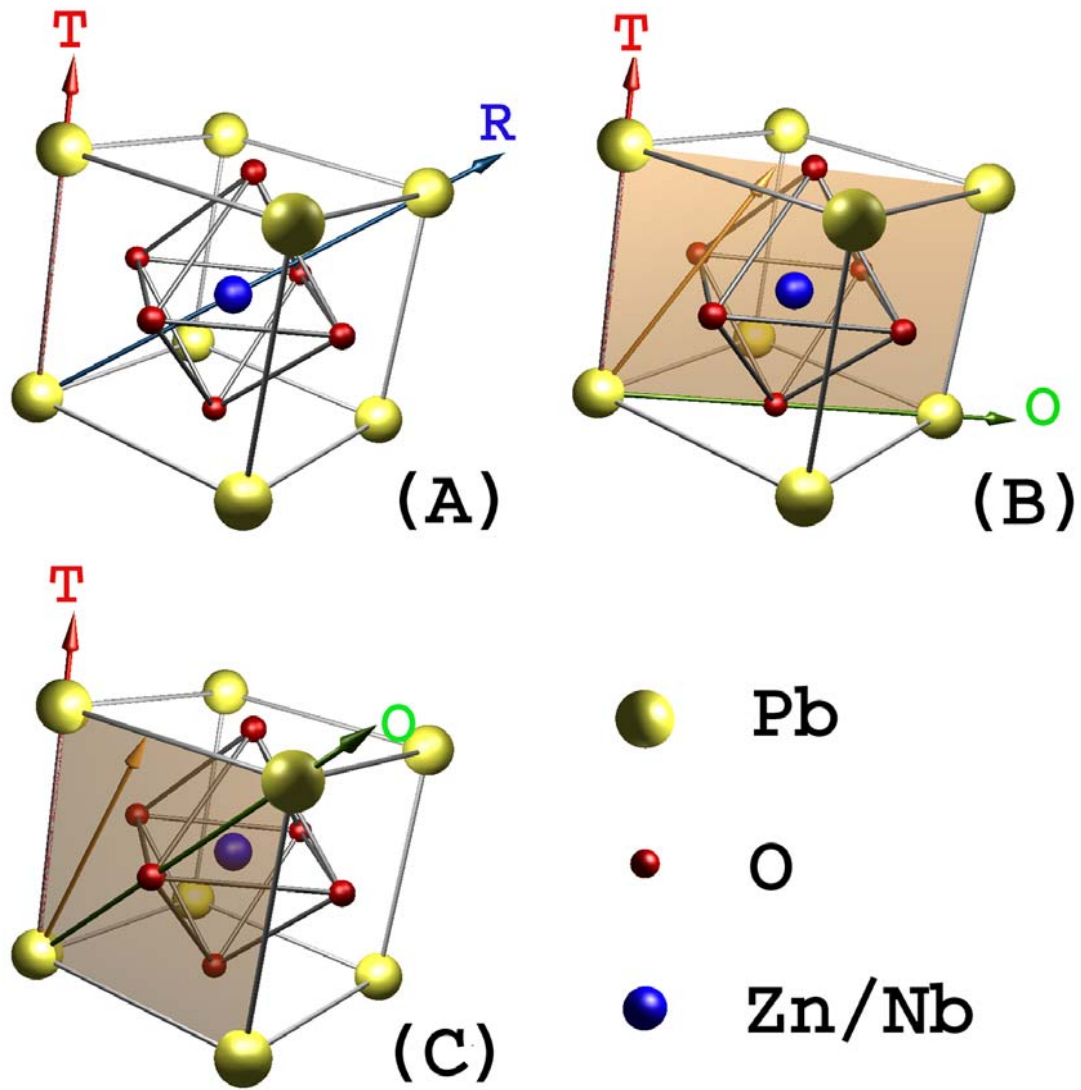

**Figure 4. Schematic diagrams of the different polarization directions (marked by arrows) in the relaxor perovskite structure.** (A) Polarizations along T and R. (B) The $M_A$ phase, for which the polarization lies in the (110) plane, can be obtained by combining T and O. (C) The $M_C$ phase, for which the polarization lies in the (100) plane,

can also be obtained by combining T and O.

**Acknowledgements** We would like to thank S. M. Shapiro and J. M. Tranquada for stimulating discussions. The financial support of the U.S. Department of Energy under contract No. DE-AC02-98CH10886, the U.S. Office of Naval Research under Grant No. N00014-99-1-0738, and the Natural Science and Research Council of Canada (NSERC) is also gratefully acknowledged.

**Competing Interests** The authors declare that they have no competing financial interests.